\documentclass{jjap3}

\usepackage{cite}
\usepackage{graphicx,color}% Include figure files
\usepackage{bm}% bold math

\makeatletter
\def\@cite#1{\textsuperscript{#1)}}
\makeatother

\newcommand{\lt}{\left}
\newcommand{\rt}{\right} %%%%%%% 

\newcommand{\kv}{{\bm k}}

\newcommand{\kf}{k_{\rm F}}

\newcommand{\qv}{{\bm q}}

\newcommand{\Rv}{{\mathbf R}}

\newcommand{\e}{\epsilon}

\newcommand{\kfi}{k_{{\rm F},i}}

\newcommand{\ve}{\varepsilon}

\title{Theory of carrier transport in graphene double-layer structure \\with carrier imbalance}

\author{Kazuhiro Hosono and Katsunori Wakabayashi}

\inst{International Center for Materials Nanoarchitectonics (WPI-MANA), 
National Institute for Materials Science (NIMS), Tsukuba, Ibaraki, 305-0044, Japan}

\abst{The carrier mobility of a graphene double-layer system is
evaluated on the basis of the Boltzmann transport theory. In this system, two
graphene layers are separated by a dielectric barrier layer. We focus on
the cases in which there is carrier imbalance between the two layers. It is
found that the mobility can be improved by controlling the carrier
density polarization between the two layers if we choose an appropriate
dielectric environment.} 

\begin{document}
\maketitle

\section{Introduction}
Graphene, a one atomic-thickness carbon sheet, has attracted much
interest owing to its unique electronic properties such as the
half-integer Hall effect and ultra -high mobility.
The electronic states of graphene near the Fermi energy are well
described by the two-dimensional massless Dirac equation. 

Recently, electronic devices composed of graphene and other
atomically thin materials have also been proposed \cite{Novoselov2012,
Kim2011b, Dean2010, Ponomarenko2011, Britnell2013}. One such system is the graphene
double-layer system, in which two graphene layers are separated by a
thin dielectric, as shown in Fig. 1(a)~\cite{Kim2011b, Dean2010, Ponomarenko2011, Britnell2013, Ryzhii2012, Ryzhii2013, Ishikawa2013}. Since the interlayer interaction
can be controlled by adjusting the interlayer distance, the GDLS has been
considered to be a good platform for studying the exciton
superfluidity~\cite{Pikalov2012, Mink2012, Abergel2013}, Coulomb drag
effect~\cite{Mink2012, Scharf2012} and plasmon mode~\cite{Profumo2012,
Badalyan2012, Stauber2012}. The Coulomb drag effect was demonstrated
using Al$_2$O$_3$\cite{Kim2011b} and h-BN~\cite{Dean2010, Ponomarenko2011, Lee2012, Zomer2012, Gorbachev2012} 
as middle dielectrics. Theoretical analysis of the device
performance of GDLS has only just begun.  

Carrier mobility is one of the key benchmarks of device performance
because it determines the power dissipation and switching speed of the device. 
Recent theory suggests improving carrier mobility by placing a
high-$\kappa$ overlayer on a semiconductor nanostructure,
which leads to the weakening of Coulomb scattering due to the screening
effect~\cite{Jena2007, Adam2007}. 
Indeed, several electronic transport measurements of graphene or
atomically -thin material have successfully revealed mobility enhancement
via change in the dielectric environment~\cite{Jang2008a, Kim2009,
Radisavljevic2011, Hollander2011, Li2013}.  
In our previous paper, we have constructed a formulation for evaluating the
dependence of the interlayer distance and the effect of the dielectric environment
on the charged-impurity-limited carrier mobility of the GDLS on the
basis of the Boltzmann transport theory.  

We have pointed out that
the carrier mobility of GDLS strongly depends on the dielectric
constant of the barrier layer when the interlayer distance becomes larger
than the inverse of the Fermi wave vector~\cite{Hosono2013}. However, we
have considered only the case in which the carrier concentrations at each
layer are equivalent ($n^{(1)}_{c}=n^{(2)}_{c}$), and have neglected
carrier density polarization, for simplicity. Since the carrier density at each layer can
 be controlled by adjusting the gate voltage in the experiments, it is necessary to develop a theory
 that takes carrier imbalance into account.   

In this study, we focus on the case in which there is carrier
imbalance. We evaluate the effect of carrier density polarization
(denoted by $\Delta n_{c}$) on the charged-impurity-limited carrier
mobility in the GDLS by extending our previously -reported
formulation. It is found that the carrier density polarization dependence of
the carrier mobility is affected by the surrounding dielectrics and interlayer
distance. The carrier mobility in the presence of carrier density
imbalance strongly depends on the interlayer distance if we set
particular constants of the dielectric environment. Our result offers appropriate ranges of the 
carrier polarization and dielectric constant of the surrounding
dielectrics to improve the charged-impurity limited mobility of GDLS. 

\section{Model and formulation}
\begin{figure}
\center
\includegraphics[width=0.5\textwidth]{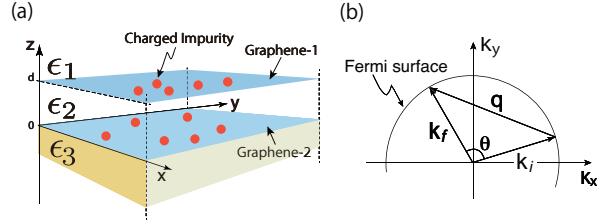}
 \caption{
(a) Schematic of GDLS with three different dielectrics.
 The interlayer distance between two graphene layers is defined as $d$. The dielectrics with $\e_1$ and $\e_3$
 are assumed to be much thicker than the interlayer distance $d$. The top and
 bottom graphene layers are numbered $1$ and $2$, respectively. The red circles
 represent randomly distributed charged impurities.
(b) Carrier scattering on a Fermi surface of radius $k_F$. An initial
state with a wave vector $\bm{k_i}$ is scattered by a charged impurity potential to
 a final state with a wavevector $\bm{k_f}$, where
 $|\bm{k_i}|=|\bm{k_f}|=k_F$. 
  Here
 $\bm{q}=\bm{k_f}-\bm{k_i}$, and $\theta$ is the scattering angle. 
} 
\label{fig:1}
\end{figure}

Figure~\ref{fig:1}(a) shows a schematic of the GDLS, in which
two graphene layers are separated by three different
dielectrics, i.e., $\e_1,\e_2$ and $\e_3$.
We assume that the two graphene layers are coupled only through 
the Coulomb interaction between the charged impurities and carriers. The Hamiltonian can be written as
\begin{align}
H=&\gamma\sum_{\kv, s, s^\prime}\sum_{i=1}^2c^{\dagger}_{\kv,s,i}\lt(\sigma_x k_x
 +\sigma_y k_y\rt)c_{\kv,s^\prime,i} 
\\
+&\frac{1}{L^2}\sum_{\kv\qv}\sum_{s,s'}\sum_{i,j} W_{ij}(q,
 d)c^{\dagger}_{\kv+\qv, s, i}c_{\kv, s', i}\rho^{(j)}_{\rm imp}(\qv), \nonumber 
\end{align}
where $c^{\dagger}_{\kv,s,i}$ ($c_{\kv,s,i}$) is the creation
(annihilation) operator for an electron with the wave vector $\bm{k}=$($k_x$,$k_y$) and
the pseudospin $s$ on the $i$-th graphene layer. Here, $\gamma = 6.46$
eV$\cdot$\AA\ is the band parameter. $\sigma_x$ and $\sigma_y$ are the
Pauli spin matrices for pseudospin; $s,s'=\pm1$ are pseudospin labels for
describing the sublattice of the honeycomb lattice. 
$L^2$ is the area of each graphene layer.  $W_{ij}$ denotes the Fourier
component of the screened Coulomb potential, which depends on the interlayer
distance $d$, and includes the effect of the Coulomb interaction
between carriers on each graphene layer through the polarization
function~\cite{Ando2006}. 
$\rho^{(j)}_{\rm imp}(\qv)=\sum^{N_{imp}}_{\alpha}e^{-i\qv\cdot \Rv^{(j)}_{\alpha}}$ is the
particle density of random impurities on the $j$-th graphene layer having
the total number of impurities $N_{imp}$. $\Rv^{(j)}_\alpha$ represents
the position of the impurities on the $j$-th layer.
The inverse of the carrier mobility $\mu^{-1}$ can be described by $\mu^{-1}=\mu^{-1}_{11}+\mu^{-1}_{22}+\mu^{-1}_{12}+\mu^{-1}_{21}$
\cite{Nomura2006, Ando2006a}. Here, $\mu^{-1}_{11}$ ($\mu^{-1}_{22}$) is the contribution of the intralayer scattering
rate of the first (second) graphene layer, and
$\mu^{-1}_{12}$ and $\mu^{-1}_{21}$ are the contributions of interlayer scattering. 
According to the semiclassical Boltzmann theory, the inverse of the each contribution of carrier mobility is given by
\begin{align}
\frac{1}{\mu_{ij}(k_{i})}=&\frac{n^{\rm imp}_{ij} \sqrt{n^{(i)}_{c} \pi}}{e\gamma} D(k_{i})\int^{\pi}_{0}d\theta \lt|W_{ij}\lt(q_i, d\rt)\rt|^2(1-\cos^2\theta),
\label{eq:mij}
\end{align}
where $n^{\rm imp}_{11} (n^{\rm imp}_{22})$ is the impurity
concentration on the first (second) graphene layer. The impurity
concentration for interlayer scattering is given as the average of
two layers $n^{\rm imp}_{12}\equiv ( n^{\rm imp}_{11}+n^{\rm
imp}_{22})/2$. 
For simplicity, we assume that the impurity concentration at each layer
$n_{\rm i}=n^{\rm imp}_{11}=n^{\rm imp}_{22}$ is equivalent
and that the Fermi level of both graphene layers lies in the conduction band. 
$D(k_i)=gk_i / 2\pi\gamma$ is the density of states of single-layer graphene 
with $g=4$ owing to the valley and spin degeneracy. 
$\theta$ is the scattering angle, and $q_i=2\kfi
\sin({\theta}/{2})$ is the scattering wave vector on the circular
two-dimensional Fermi surface, as shown in Fig.~\ref{fig:1}(b). The Fermi wave number on each
graphene layer is given as $\kfi=\sqrt{4\pi n^{(i)}_{c}/g}$. 
Note that the last $\theta$-dependent factor also contains the phase of the wave
function of graphene \cite{Ando2006,Nomura2006}. 
The structural parameter such as the interlayer distance $d$ and
scattering potential due to charged impurities are included in the
screened potentials $W_{ij}$. 

Here, we briefly explain the derivation of the screened Coulomb potential
$W_{ij}$ from
the unscreened one $v_{ij}$. The analytical expression of the unscreened Coulomb
potentials of the GDLS can be derived
using the image charge method~\cite{Kumagai1989, Jena2007, Li2013}. 
For this system, we need to consider an infinite series of point image charges
arising from two interfaces at $z = 0$ and $d$ shown in Fig. 1(a), where two types of
dielectrics are spanned by a graphene layer. 
The resulting unscreened Coulomb potentials are given as
\begin{align}
v_{11}(q_1, d)=&\frac{4\pi e^2}{q_1}\frac{\e_2+\e_3\tanh(q_1d)}{X_1},
\label{eq:bare1}
\\
v_{22}(q_2, d) =&\frac{4\pi e^2}{q_2}\frac{\e_2+\e_1\tanh(q_2d)}{X_2},
\label{eq:bare2}
\\
v_{12}(q_1, d)=&\frac{4\pi e^2}{q_1}\frac{\e_2}{X_1 \cosh(q_1d)},
\label{eq:bare3}
 \\
v_{21}(q_2, d)=&\frac{4\pi e^2}{q_2}\frac{\e_2}{X_2 \cosh(q_2d)}.
\label{eq:bare4}
\end{align}
Here $v_{11}$ and $v_{22}$ are the intralayer Coulomb interactions
on the first and second graphene layers, respectively. $v_{12}$ and $v_{21}$ are the interlayer Coulomb
interactions, and we define
\begin{align}
X_i=&\e_2(\e_1+\e_3)+(\e_2^2+\e_1\e_3)\tanh(q_id).
\end{align}
 
The above potentials have been used in the context of the superfluid
magnetoexitons \cite{Pikalov2012} and plasmon mode \cite{Badalyan2012,
Profumo2012} of the GDLS. 
The above expressions indicate that the parameter $q_id$ ($\approx \kf d$) determines the
screening behavior and the strength of the interlayer Coulomb interaction. 

The screened Coulomb potentials are described by the random phase
approximation (RPA) as
\begin{align}
W=&V+V\Pi W
=V(1-V\Pi )^{-1},
\end{align}
where 
\[
 V=
  \begin{pmatrix}
   v_{11} & v_{12} \\
   v_{21} & v_{22}
  \end{pmatrix},
\]

\[
 \Pi=
  \begin{pmatrix}
   \Pi_{1} & 0 \\
   0 & \Pi_{2}
  \end{pmatrix}.
\]
Here, $\Pi_{1}  (\Pi_{2})$ is the static
polarization function of the first (second) graphene layer \cite{Ando2006a, Nomura2006}.
The static polarization is written as
\begin{align}
\Pi_i=\Pi^{(+)}_i+\Pi^{(-)}_i,
\end{align}
where $\Pi^{(+)}_i$ and $\Pi^{(-)}_i$ are given as
\[
\Pi^{(+)}_i=
  \begin{cases}
  D(\kfi) \lt(1-\frac{\pi}{4}\frac{q_i}{2\kfi}\rt) & (\frac{q_i}{2}\le\kfi)
   \\
  D(\kfi) \lt(1-\frac{1}{2}\sqrt{1-4(\frac{\kfi}{q_i})^2}-\frac{1}{4}\frac{q_i}{\kfi}\arcsin\lt(\frac{2\kfi}{q_i}\rt)\rt)  & (\frac{q_i}{2}> \kfi),
  \end{cases}
\]
\begin{align}
\Pi^{(-)}_{i}=&D(\kfi)\frac{\pi}{8\kfi}q_i.
\end{align}
From Eq. (8), the effective interactions are obtained as
\begin{align}
W_{11}=&\frac{1}{\ve}\lt(v_{11}+(v_{11}v_{22}-v_{12}v_{21})\Pi_{2}\rt),
\label{eq:11}
\\
W_{22}=&\frac{1}{\ve}\lt(v_{22}+(v_{11}v_{22}-v_{12}v_{21})\Pi_{1}\rt),
\label{eq:22}
\end{align}
\begin{align}
W_{12}=&\frac{v_{12}}{\ve},
\label{eq:12}
\\
W_{21}=&\frac{v_{21}}{\ve}.
\label{eq:21}
\end{align}
where the intralayer interactions $v_{11}$ and $v_{22}$, and the interlayer interactions $v_{12}$ and $v_{21}$ are defined in Eqs. (3)-(6), respectively.
Here, the dielectric function is defined as 
\begin{align}
\ve(q_1,q_2)=&\det(1-V\Pi)=(1+v_{11}\Pi_{1})(1+v_{22}\Pi_{2})-v_{12}v_{21}\Pi_{1}\Pi_{2}.
\label{eq:ve}
\end{align}

By combining Eqs. (\ref{eq:11})-(\ref{eq:ve}),
we can evaluate the carrier mobility of GDLS in the presence of an
imbalance in the concentration between two graphene layers, i.e., $n^{(1)}_c\neq n^{(2)}_c$. 

\begin{figure*}[t]
\center
\includegraphics[width=0.95\textwidth]{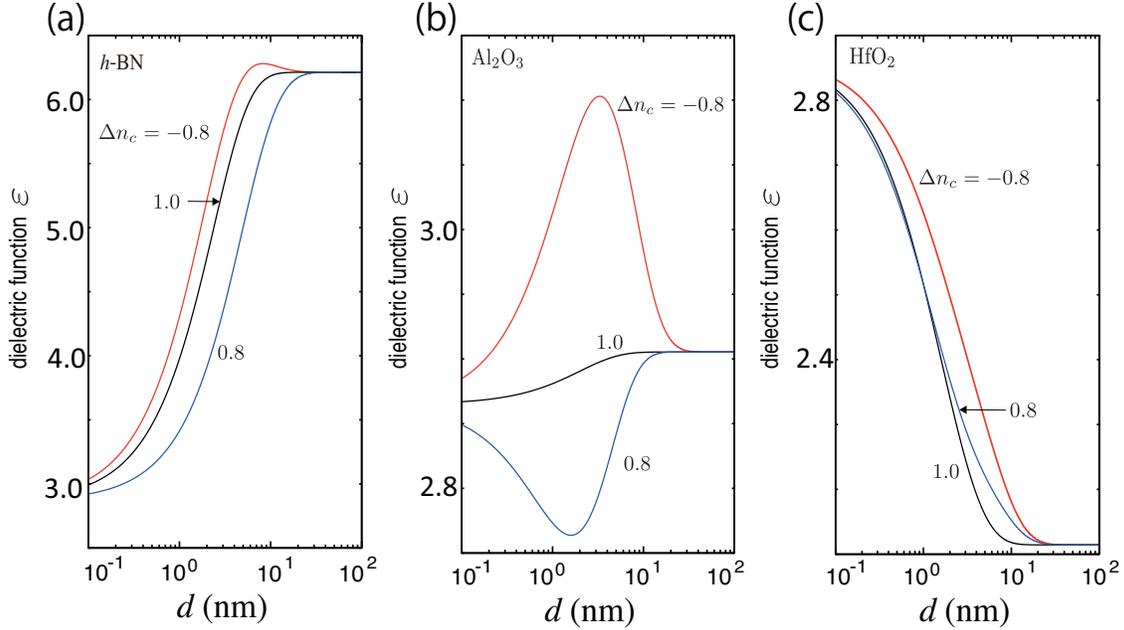}
 \caption{Interlayer distance $d$ dependence of the dielectric function at the scattering angle $\theta=\pi/2$ for three different 
 density polarization $\Delta n_{c}=-0.8,1$ and $0.8$. Left, middle, and right panels represent the cases of h-BN, Al$_2$O$_3$ and HfO$_2$, respectively. Here the total carrier density is $n^{(1)}_{c}+n^{(2)}_{c}=2 \times 10^{12}/{\rm cm^2}$.}   
 \label{fig2}
 \end{figure*}

\begin{figure*}[t]
\center
\includegraphics[width=0.95\textwidth]{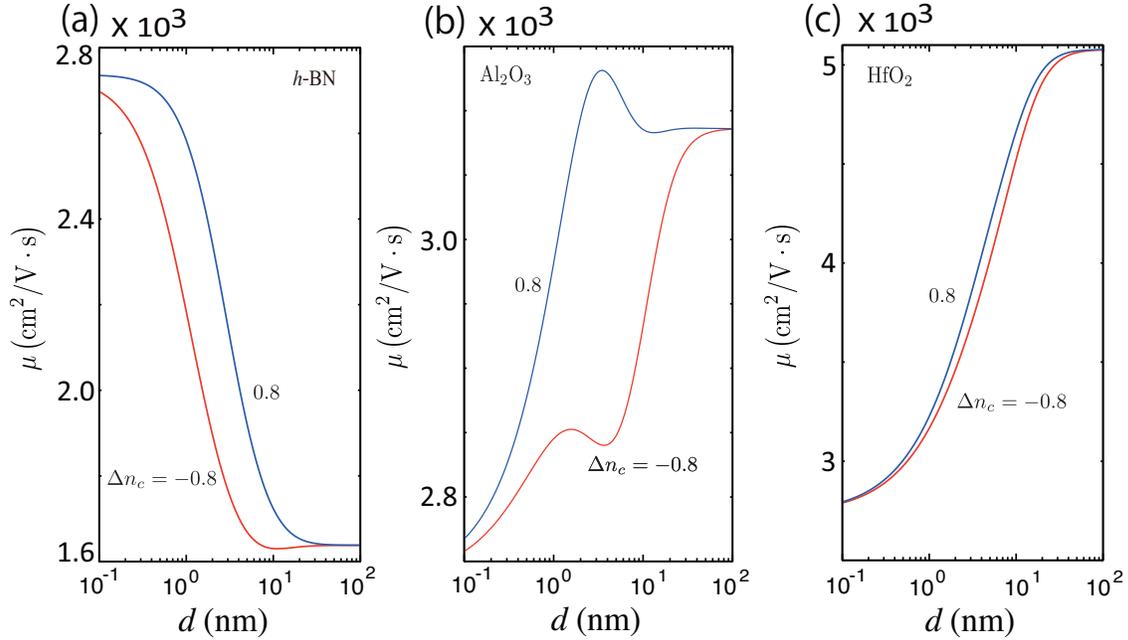}
 \caption{(a) Plots of mobility versus interlayer distance for different carrier
 density polarizations, $\Delta n_{c}=-0.8,0.8$. Here, the total carrier
 density is $n^{(1)}_{c}+n^{(2)}_{c}=2 \times 10^{12}/{\rm cm^2}$, the
 impurity density is $n^{(1)}_i=n^{(2)}_i=5 \times 10^{11}/{\rm cm^2}$,
 $\e_1=\e_{\rm Air}=1$, and $\e_2=\e_3=\e_{\rm Al_2O_3}=12.53$.}
 \label{mnc-fig3}
 \end{figure*}

\begin{figure*}[t]
\center
\includegraphics[width=0.475\textwidth]{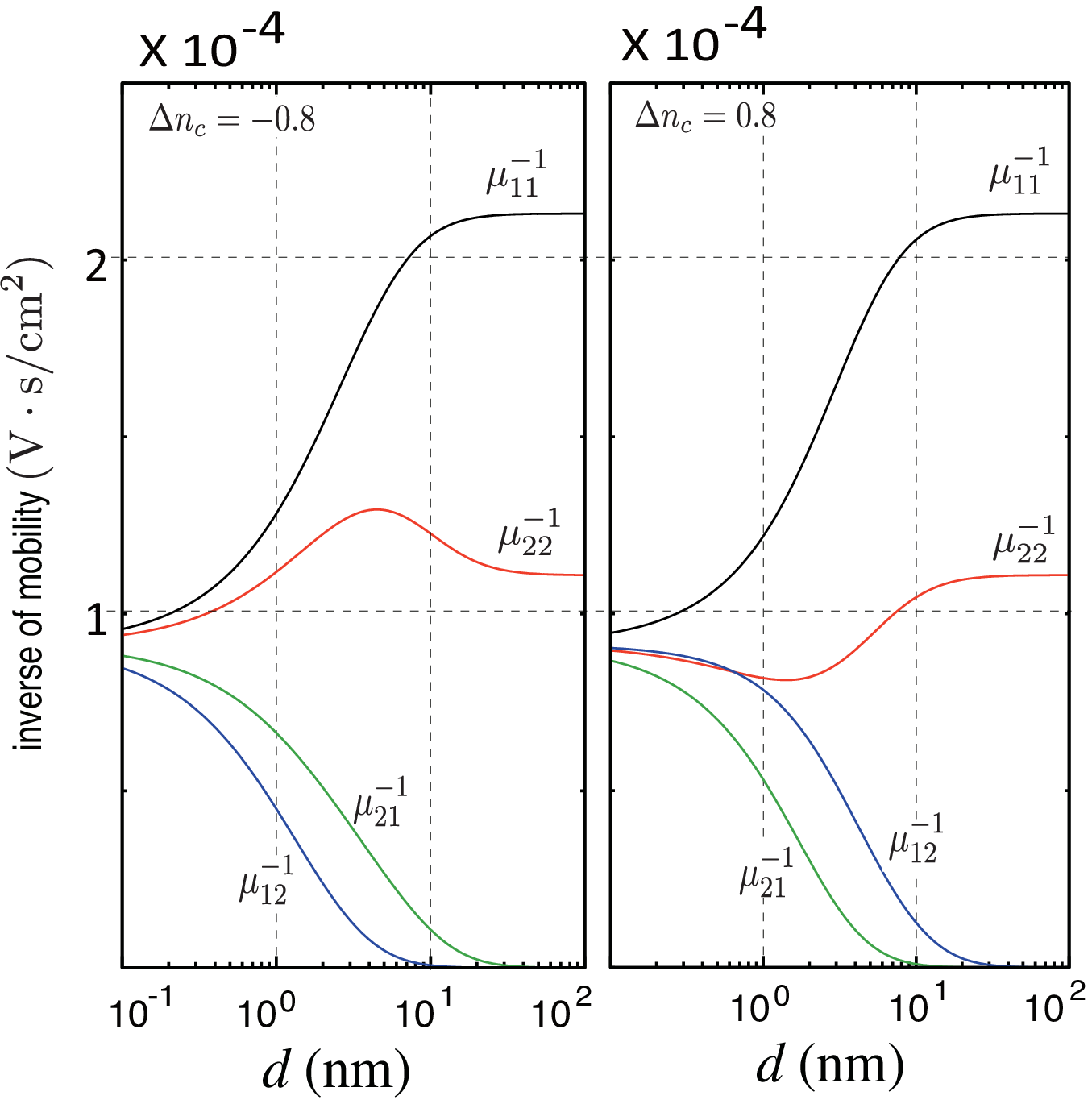}
 \caption{ Four components of the inverse of mobility as function of the interlayer
 distance for $\Delta n_{c}=-0.8$ (left) and $\Delta n_{c}=0.8$
 (right). 
}
 \label{mnc-fig4}
 \end{figure*}

\section{Results and discussion}
We first investigated the dependence of the dielectric function
Eq. (\ref{eq:ve}) on the interlayer distance $d$ and carrier
density polarization. Here, we define the carrier density polarization
as $\Delta n_{c}=\lt(n^{(2)}_{c}-n^{(1)}_{c}\rt)/\lt(n^{(1)}_{c}+n^{(2)}_{c}\rt)$, and fix
the total carrier concentration at $n^{(1)}_{c}+n^{(2)}_{c}=2 \times
10^{12}/{\rm cm^2}$. When the carriers are only in the first (second)
layer, $\Delta n_{c}=-1(1)$. In the case of $\Delta n_{c}=0$ in which
the two layers have identical carrier densities, the dielectric function
captures the screening effect of scattering potentials due to
charged impurities. In order to see the role of the middle dielectric layer, 
we assume the dielectric constants for the top and bottom layers as $\e_{1}=\e_{\rm Air}=1$ and $\e_{3}=\e_{\rm Al_20_3}=12.53$, respectively, and consider three different dielectrics as the middle layer,
i.e., h-BN, Al$_2$O$_3$, and ${\rm HfO_2}$. Their
dielectric constants are $\e_{\rm h-BN}=4$, $\e_{\rm Al_2O_3}=12.53$, 
and $\e_{\rm HfO_2}=22$, respectively\cite{Frederisk1991, Desgreniers1991, Kang2000}. 
 
Figure 2 shows the $d$ dependences of the dielectric function at
a scattering angle $\theta=\pi/2$ for several different density
polarizations, i.e., $\Delta n_{c}=-0.8, 1,$ and $0.8$ for three different
middle dielectrics. Figures 2(a)-(c) represent the
cases of h-BN, Al$_2$O$_3$ and HfO$_2$,
respectively. We can see that screening effect at $\theta=\pi/2$ 
is enhanced with increasing interlayer distance in the case of h-BN ($\e_2 < \e_3$), as shown in Fig. 2(a), but
reduced for HfO$_2$ ($\e_2 > \e_3$), as shown in Fig. 2(c). We see that the $\Delta n_{c}$
dependences in these two cases are weak, because of the large mismatch
between $\e_2$ and $\e_3$. However, for Al$_2$O$_3$ ($\e_2 = \e_3$), we can see that the presence of carrier polarization leads to very different behaviors of the interlayer distance as shown in Fig. 2(b), together with a peak structure $\Delta n_{c}=-0.8$. However, at $\Delta n_{c}=0.8$, a dip appears instead. Since the $d$ dependence of the dielectric screening effect in the case of $\e_2 \simeq \e_3$ is weaker than those in other cases, the effect of the carrier density polarization becomes prominent.

 %Here we choThe interlayer distance dependence of dielectric function
 %strongly depends on the middle dielectrics.   %w 
%The $\Delta n_c$ dependence of the dielectric function of middle panel
%in Fig. 2 affects to the carrier mobility.  
Figures 3(a)-(c) show the dependences of the total mobility at the
interlayer distance $d$ for two different carrier polarizations $\Delta n_{c}=-0.8$ and $0.8$ in the same dielectric environment as in Figs. 2(a)-(c), respectively. These mobilities are strongly affected by dielectric functions. In particular, when we choose $\rm Al_2O_3$ as the middle dielectric, the mobility in Fig. 3(b) shows a dip for $\Delta n_c=-0.8$ and a peak for $\Delta n_c=0.8$, arising from the peak or dip structure in the dielectric functions, as shown in Fig. 2(b). 

We show, in Fig. 4, the $d$ dependence of the four
components of the inverse of the mobility at $\Delta n_{c}=-0.8$ (left)
and $0.8$ (right). By comparing the left and right panels in
Fig. 4, we find that the interlayer components $\mu_{12}$ and
$\mu_{21}$ are exchanged by changing the carrier polarization.  
The intralayer component $\mu_{11}$ is almost unaffected by the
inversion of the carrier polarization, because the
$\Delta n_{c}$ dependence is canceled out
in the effective potential $W_{11}$ [Eq. (\ref{eq:11})] for
$\epsilon_1\ll \epsilon_2, \epsilon_3$. 

On the other hand, another intralayer component, $\mu_{22}$, considerably
depends on the interlayer distance $d$ and carrier density polarization.
The reason for this is that the $\Delta n_{c}$ dependence of the dielectric
function cannot be canceled out in the effective potential [Eq. (\ref{eq:22})].  
We found that the characteristic $d$ and $\Delta n_c$ dependences of the carrier mobility
for $\e_1 \ll \e_2 \simeq \e_3$ were dominated by one of the intralayer components, i.e., $\mu_{22}$.  
 
In summary, we have investigated the carrier transport of GDLS in the
presence of carrier polarization by extending our previous theory. We showed 
that the carrier mobility considerably depends on the carrier density polarization 
when the dielectric environment parameters $\e_1, \e_2,$ and $\e_3$ are
 satisfied under the condition $\e_1 \ll \e_2 \simeq \e_3$. Our result reveals that the mobility can be improved
by choosing higher-dielectric-constant materials as well as introducing the carrier polarization between two layers.
From our results, we proposes guidelines for
experiments  on and applications of new functional atomically thin devices. 

\acknowledgment
This work is supported by Grants-in-Aid for Scientific Research (KAKENHI)
(Nos. 25107005, 25107001, 224710153, and 23310083) from the Japan Society for the Promotion of Science.

\end{document}